\documentstyle[11pt,emulateapj,epsfig,psfig]{article} %<===ApJformat
\lefthead{Nobili et al.}
\righthead{PHASE LAGS IN GRS 1915+105}

\begin{document}

\title{A Comptonization Model for Phase Lag Variability in GRS 1915+105}

\author{L. Nobili, R. Turolla, L. Zampieri}
\affil{Department of Physics, University of Padova, \\
Via Marzolo 8, I--35131 Padova, Italy \\
e--mail: nobili@pd.infn.it, turolla@pd.infn.it, zampieri@pd.infn.it}
\and
\author{T. Belloni}
\affil{Osservatorio Astronomico di Brera, via Bianchi 46, I--23807 Merate,
Italy \\e--mail: belloni@merate.mi.astro.it}

\begin{abstract}

In this {\it Letter\/} we propose a simple thermal comptonization model to
account for the observed properties of the phase lags associated to the
``plateau'' intervals of GRS 1915+105. By invoking a temperature
stratification in a corona and assuming that the optical depth of the
comptonizing region increases as the disk inner radius moves inward, we
are able to reproduce both the observed colors and time lags in the
continuum.

\end{abstract}


\section{Introduction}\label{sec-intro}

The galactic microquasar GRS 1915+105, discovered in
1992 by GRANAT (Castro-Tirado, Brandt \& Lund \cite{castro:1992}), is a transient
source with an extremely wide variety of variability
modes (Greiner et al. \cite{greiner:1996}; Belloni et al. \cite{belloni:1997}; Muno, Morgan
\& Remillard \cite{muno:1999}).
During the period 1996-1997 it has been extensively
and repeatedly observed by the Rossi X-ray Timing
Explorer (RXTE), alternating phases of dramatic
variability to phases of remarkably regular behavior
(Belloni et al. \cite{belloni:2000}).

GRS 1915+105 was the first galactic source to
show superluminal radio expansion (Mirabel \& Rodr\`\i guez \cite{mirabel:1994}),
usually interpreted in terms of emission from gas
expanding at relativistic velocities.
Measurements of this superluminal motion
allow to estimate the source distance, which is
in the range 6--12.5 kpc (Rodr\`\i guez \&
Mirabel \cite{rodrig:1999}; Fender et al. \cite{fender:1999}).
Because of the high extinction along the line of
sight, only a faint infrared counterpart has been discovered (Mirabel et al.
\cite{mirabeletal:1994}).
As a consequence, neither the mass function nor
the orbital period are known. The source is believed
to host a black hole because it shows properties
(variability and superluminal motion) very
similar to the other known galactic microquasar,
GRO 1655-40, for which a precise dynamical
measurement yields $M\sim 7 M_\odot$ (Orosz \& Bailyn \cite{oro:1997}).

Quasi Periodic Oscillations (QPOs) have been
observed in GRS 1915+105 in an incredibly large interval of
frequencies, ranging from about 0.001 to 67 Hz
(Morgan, Remillard \& Greiner \cite{morgan:1997};
Chen, Swank \& Taam \cite{chen:1997}). In particular, in
some observations a QPO with variable
centroid frequency ($\nu_{QPO}\sim 0.5$--10 Hz), whose value
and root mean square amplitude correlate
with the thermal flux, is detected (Markwardt, Swank \& Taam
\cite{mark:1999}; Trudolyubov, Churazov \& Gilfanov \cite{trudo:1999}).

The X-ray spectrum of GRS 1915+105 shows all the distinctive
properties of other Galactic
Black Hole Candidates (BHCs). A power-law tail
with variable luminosity and slope is present
at high energies, while a thermal component often
dominates at low energies (Belloni et al. \cite{belloni:1997};
Muno, Morgan \& Remillard \cite{muno:1999}). A detailed spectral
classification of the different states of GRS
1915+105 has been recently presented by
Belloni et al. (\cite{belloni:2000}).
According to these authors, the variability of the X-ray emission is associated
with the alternation of three basic states, called A, B and C (see also Markwardt, Swank
\& Taam \cite{mark:1999}; Muno, Morgan \& Remillard \cite{muno:1999}). During states A and B,
corresponding to relatively
high and soft flux, a strong thermal disk component dominates the emission. In
state C, the 3--25 keV flux is characterized by lower luminosity and harder X-ray
colors, being dominated by a hard ($\Gamma=1.5$--2.5) power law with much softer (or
even unobservable) thermal component. State C can last uninterruptedly,
for a long
time (weeks to months, the so called plateau intervals), in which case no soft component is observed at all. State C is
always associated with the 0.5--10 Hz QPO and a flat-top component in the Power
Density Spectrum (Markwart, Swank \& Taam \cite{mark:1999}; 
Trudolyubov, Churazov \&
Gilfanov \cite{trudo:1999}; Reig et al. \cite{reig:2000}).

The analysis of a set of
RXTE observations of GRS 1915+105 in this state showed that phase lags are
present, both in the continuum and the QPO (Reig et al. \cite{reig:2000}).
The sign of the lag is
correlated with QPO frequency, count rate and X-ray colors.
The lag is positive (photons in the 5--13 keV band
arrive later than those in the 2--5 keV band) when
the spectrum is hard and the centroid of the
0.5--10 Hz QPO is below 2 Hz; it then becomes negative (soft
photons are delayed) as the spectrum softens and
the QPO centroid frequency increases above 2 Hz.
The smooth correlations of the lags with colors suggests that a unique
mechanism is driving the observed transition from positive to negative lags. 

In this letter we propose a simple comptonization
model to account for the
observed properties of GRS 1915+105 in state C, which is
outlined in \S~\ref{sec-model}. In \S~\ref{sec-results} we
report the main results obtained and compare them with
observations. Finally, \S~\ref{sec-discuss} contains a
short discussion of the implications of our results.

\section{The Model}\label{sec-model}

No definite model for the accretion
flow in GRS 1915+105 has been presented as yet. Here we consider a highly idealized
scenario in which a Shakura-Sunyaev disk
(Shakura \& Sunyaev \cite{shaksuny:1973}) coexists with a hotter component.
As suggested by Belloni et al.
(\cite{belloni:1997}), the main cause for the
large-scale time variability of this source is
the onset of a thermal-viscous instability which blows off the inner,
radiation-pressure supported region of the disk.
In particular,
according to this scenario, the hard state corresponds
to the replenishment by the disk of the
previously evacuated region on a viscous timescale.
Further support to this picture has been provided by numerical simulations
(Szuszkiewicz \& Miller \cite{ewa-john:1997}; \cite{ewa-john:1998}).

Fitting the observed spectra of BHCs in the hard state requires the
presence, in addition to a standard disk, of a hot, less dense
phase, formed e.g. by the debris of the
puffed-up inner region or by the evaporation of the disk itself.
The upscattering of soft disk photons by a hot
cloud at $T\approx 10$ keV can account also for the
positive lags of the continuum. In fact, Compton cloud models at constant
temperature
have been already proposed in
connection with other sources which show (positive) time delays
(Miyamoto et al. \cite{miyamo:1991}; Vaughan et al.
\cite{vaug:1994}; see also Poutanen \cite{put:2000} for a
review). Compton models suffer from the problem that
soft photons cool down the cloud to the temperature
of the impinging radiation, unless an
efficient heating mechanism is at work. However,
if cooling of ions is inefficient, like in advection-dominated flows,
protons
decouple from electrons and stay hot at a fraction of their virial temperature,
$T_p(r)\sim \eta GMm_p/r$.
The electron thermal balance is fixed by the competition
between
ion heating and Compton cooling. A simple estimate of the
equilibrium electron temperature gives
\begin{equation}\label{telectron}
T(r) \approx 10 \, (c^2 r_{in}/GM)^{2/5}(1-r_{in}/r)^{-2/5} \ {\rm keV}
\end{equation}
for a scattering depth $\gtrsim 1$, $\eta =0.01$, and
$L_{soft}\propto (1/r_{in} - 1/r)$.
As expected (see Poutanen \cite{put:2000}), the cloud temperature 
diminishes as the
disk rim moves in and $L_{soft}$ increases,
but the dependence is weak. In an ion-heated corona the temperature is 
larger at smaller radii because $L_{soft}$ decreases with $r$ so that 
less photons are available to cool the inner part of the cloud.

A non-homogeneous Compton cloud model may indeed explain
the positive/negative lags and the decrease in the hard
color observed in the state C of GRS 1915+105, provided that
the corona becomes denser as the disk extends inwards. 
The density may increase in time if a cloud of fixed mass
contracts as the result of angular momentum losses caused by viscous and/or
radiation stresses.
In order to illustrate the basic idea of the model,
we sketch the corona as formed
by two regions (referred to as ``warm'' and ``hot''), each at uniform density
and temperature, $T_W$ and $T_H$, with $T_H>
T_W$ (see Figure \ref{artist}). The scattering depths
are $\tau_W$ and $\tau_H$, $\tau_H>\tau_W$. When the density
of the corona is low, soft photons get
preferentially scattered in the inner part of the corona which has 
$\tau_H\gtrsim 1$ (while
$\tau_W\lesssim 1$), gaining energy and producing a positive lag. As density increases,
the Compton $y$-parameter of the corona exceeds unity below a certain radius $r_{sc}$.
Photons emitted by the inner part of the disk undergo repeated
scatterings
with the hot electrons and start to fill the Wien peak at $T_H$.
The spectrum emitted by the disk plus the hot, dense part of the corona is therefore the
superposition of a multitemperature blackbody and a Bose-Einstein distribution
at $T_H$; the latter has the same number of photons originally emitted by the disk in
the region $r_{in}\lesssim r \lesssim r_{sc}$.
These photons have still to traverse the warm part of the corona
before escaping to infinity. In doing so, they are downscattered and this produces
the negative lag and the observed spectrum. A somehow similar geometry (a
standard disk with variable radius and a hotter, comptonizing cloud), although 
in a different context, has been
successfully invoked to explain the spectral properties of Nova Muscae 1991
(e.g. Misra \& Melia \cite{mimel:1997}; Esin, McClintock \& Narayan 
\cite{esmcnar:1997}).

In the next section we show that quantitative estimates, based on this simple picture,
are indeed in agreement with the observed colors and lags for reasonable values of the
parameters. In our model the inner radius changes only by a
factor $\sim 3$,
so, for the sake of simplicity, in the following we neglect the small ($\sim 1.5$, see
eq. [\ref{telectron}]) change of $T$ with $r_{in}$. 

\section{Numerical Estimates}\label{sec-results}

The calculation of the emerging spectrum proceeds in two steps and follows the standard
procedure discussed by Sunyaev \& Titarchuk (\cite{suntit:1980}). First, we compute the
time-dependent comptonized spectrum for a given seed photon distribution by solving
the Kompaneets equation in a semi-infinite homogeneous medium and then use an escape
probability algorithm to obtain the time evolution of the spectrum emerging from a
medium of finite depth.
%The comptonizing region has temperature $T$ and scattering depth $\tau$.
We introduce a dimensionless time $u=(n_e\sigma_T c \theta)t$ and energy
$x=h\nu/kT$, where $n_e$ is the electron density, $\sigma_T$ the Thomson
cross-section and $\theta = kT/m_ec^2$. In the following all lengths are expressed in
units of $GM/c^2$.
The photon occupation number $n(u,x)$ is the solution of the Kompaneets equation
with an appropriate initial condition $n(0,x)$.
%which follows from the assumptions made in \S~\ref{sec-model}.
For $r_{in}> r_{sc}$ the initial photon
occupation number is that of a multitemperature
blackbody (calculated including GR and hardening effects, e.g. Novikov \& Thorne
\cite{notho:1973}; Shimura \& Takahara \cite{giap:1995}) truncated at $r_{in}$ whereas,
for $r_{in} < r_{sc}$, it corresponds to the
superposition of a Bose-Einstein distribution
at $T_H$ and a multitemperature blackbody truncated at $r_{sc}$.

Since in our model the long-term evolution corresponds to a variation of the inner
edge of the disk, the Kompaneets equation has been solved numerically for
a sequence of values of $r_{in}$.
% in the range $[2\,, 30]$, the actual lower
%limit depending on the black hole angular momentum $a$.
Once $n(u,x)$ has been evaluated, the time evolution of
the escaping photons is given by
\begin{equation}\label{esc-prob}
N(u,x) = x^2n(u,x)P(u,\tau)
\end{equation}
where $N$ is proportional to the count rate and $P(u,\tau)$ is the escape
probability distribution.
The explicit form of $P(u,\tau)$ depends on the geometry (Sunyaev \& Titarchuk
\cite{suntit:1980}). In our calculations we have used a
weighted sum of the expressions for a spherical cloud with either
a central or a diffused photon source.

The total count is proportional to the convolution
\begin{equation}\label{spec-fin}
F(x) = \int_0^\infty  x^2n(u,x)P(u,\tau)\, du\, .
\end{equation}
We note that eq. (\ref{spec-fin}) gives exactly the spectrum produced
by a stationary source of photons. Therefore it will be used to compute the
hardness ratios on time intervals substantially larger than typical variability
timescales in the continuum.
On the other hand, if one is interested in following the
time evolution of the spectrum over timescales $\sim 1/\nu_{QPO}$ the relevant
quantity to consider is $N(u,x)$.

To reproduce the observed time evolution of the color in a given energy band
we have convolved $N(u,x)$
with the interstellar absorption and with the detector response function
\begin{equation}\label{eq-color}
C(E_1-E_2) \propto\int_{E_1}^{E_2} N(u,x)\exp{[-N_H\sigma(x)]}A(x)\, dx
\end{equation}
where $N_H$ is the column density.
The lags in the continuum are computed comparing the time evolution of the
the soft and hard colors, $C(2-5)$ and $C(5-13)$, as obtained from eq. (\ref{eq-color}).
The hardness ratios HR1 and HR2, defined as
$C(5-13)/C(2-5)$ and $C(13-60)/C(2-5)$, are calculated again starting from
eq. (\ref{eq-color}) but this time replacing $N(u,x)$ with $F(x)$.

The calculations presented here
have been obtained for $T_H=15$ keV, $T_W=1.5$ keV and $r_{sc}=20$; the black hole
mass and angular momentum are $M=6M_\odot$ and $a=0.1M$. The model with
$r_{in}=16$ corresponds to the usual picture in which a hot corona ($T=T_H$)
of moderate depth ($\tau=\tau_H\sim 3$) upscatters the soft photons produced by the disk;
under these conditions the energy exchange in the warm region can be reasonably assumed to be
negligible.
The opposite situation is represented by the model with $r_{in}=6$. Now the disk
inner region lies entirely within the hot part of the corona which, because of the increase
in the density, has $\tau_H\gtrsim 100$. Escaping photons are then reprocessed in the outer,
warm
part of the corona ($T=T_W$, $\tau=\tau_W\sim 10$). Intermediate cases are
more difficult to
model in the present framework, because photons scatter in both regions and
this would require a treatment of comptonization in a
non-uniform medium. To avoid undue complications, at this stage we assume that
the effects of a temperature profile can be roughly described in terms of an ``effective''
temperature $T$, $T_W<T<T_H$. For these models $T$ and $\tau$ were chosen in such a
way to maintain $y\sim 1$.

Our results are illustrated in Figure~\ref{plots}.  As
shown in Figure~\ref{plots}a, the observed hardness
ratios are quite well reproduced by the model for a
value of the column density, $N_H=2\times 10^{22}$
cm$^{-2}$, consistent with observations (Belloni et al. 
\cite{belloni:2000}). When the
inner edge of the disk is far out, the spectrum is a
power-law produced by thermal comptonization of soft
disk photons. As $r_{in}$ decreases, downscattering
becomes progressively more important with respect to
upscattering in the hot corona, the spectrum
softens and, at the same time, the phase lag of the
continuum decreases. Eventually, systematic
downscattering dominates and soft (2--5 keV) photons,
which underwent more scatterings, escape later than
hard (5--13 keV)  photons, {\it causing an inversion
in the sign of the lag\/}.  Figures~\ref{plots}b
and~\ref{plots}c show the lag of the continuum as a
function of  HR1 and QPO frequency (the scattering of the computed points is
partially due to the poor sampling of the parameter space).
Time lags have been computed assuming that the typical
size of the comptonizing corona varies with time in
accordance with $r_{in}$ in the range $\sim$20--400. The relation between
$\nu_{QPO}$ and $r_{in}$ was
derived associating the
model with $r_{in}= 6$ (16) to the observation with
the largest (smallest) QPO frequency, 6 (0.6) Hz and yields
\begin{equation}
r_{in}\simeq 13\nu_{QPO}^{-0.43} \,.
\label{rinu}
\end{equation}

\section{Discussion}\label{sec-discuss}

In this {\it Letter\/} we have shown that a simple
Compton model can account for the main observational
properties of GRS 1915+105 in state C. In particular,
the fundamental correlations (color-color, lag-color
and lag-QPO frequency) are successfully reproduced
for a reasonable choice of the model parameters.
Although we have adopted here a quite definite picture for
the accretion flow, it is important to stress that the
basic results are largely independent of the specific
choice for the geometry. The decrease of the lag in
the continuum and the softening of the spectrum as the
disk fills in are simply a consequence of the increase
of the coronal depth and of the decrease of the
temperature where radiation scatters before escaping.
Thus, only three main ingredients are required to
account for the observed correlations within the
framework of a simple thermal comptonization model: a
non-isothermal corona of varying optical depth, a
truncated standard disk and a source of more energetic
($\approx 10$ keV) primary photons at small radii. Further
assumptions (like the variable size of the corona and
the existence of an ``effective'' comptonization
temperature) are introduced only to match the observed
values. It should be noted, however, that other
geometrical settings for the time-varying accretion
flow may work equally well.

On the other hand, albeit not directly supported by
hydrodynamics computations, the flow structure adopted
here exhibits a certain degree of consistency in the
way in which the QPO frequency and the coronal optical
depth vary with inner disk radius. In fact, without
entering into the mechanism that originates the
observed QPO, we have assumed that it is somehow
produced at the inner edge of the disk and that its
frequency varies with $r_{in}$. It is interesting to note that
the relation between $\nu_{QPO}$ and $r_{in}$ derived in \S~\ref{sec-results}
(eq.~[\ref{rinu}]) is similar
to that found by
Di Matteo \& Psaltis (\cite{dima:1999}) combining the
empirical relation of Psaltis, Belloni \& van der Klis
(\cite{psaltis:1999})  with the upper bound provided
by the keplerian orbital frequency. We stress,
however, that our result has been obtained in an
independent way and is a direct consequence of the
values of the inner disk radius needed to match
the observed colors and time lags. The contraction of the
corona (a factor $\sim 20$) required to reproduce the
largest/smallest lags implies an increase in the density ($\rho\propto
R^{-3}$ at constant
coronal mass) consistent with the assumed ratio of the scattering depths,
$\tau_H(r_{in}=6)/\tau_H(r_{in}=16)\sim 100$.
The smallest value of
the inner disk radius required by the model, $r_{in}
\simeq 6 M$, is consistent with the assumed value $a/M=0.1$.
If the black hole is rapidly rotating, as
claimed by Zhang, Cui \& Chen (\cite{zhang:1997})  and
Cui, Zhang \& Chen (\cite{cui:1998}), then in low
luminosity state C  the disk does not
reach the innermost stable circular orbit.

A complete discussion of the behaviour of GRS 1915+105
in states A and B is outside the scope of the present
investigation. Here, we simply mention that models with
high coronal optical depth but hard-photon starved (because, e.g.,
of the final collapse of the innermost, denser part
of the corona) have hardness ratios in agreement with
those observed in states A and B. Although no QPO has been
observed in these states as yet, the positive detection of time lags
correlated with colors in states A and B may lend
further support to the present Compton model.

Very recently, complex phase lag behavior in 
XTE J1550-564 has been reported by Cui, Zhang \& Chen (\cite{czc:2000}). 
The power spectra are very similar to those in Reig et al. (\cite{reig:2000}).
However, the QPO phase-lag dependence is different. The fundamental/first 
harmonic shows a negative/zero lag,
increasing in magnitude as the spectrum softens.
In the light of the higher complexity of the timing
properties of XTE J1550-564, any attempt to extend the present model to 
this source is premature.

%\acknowledgments
%Work partially supported by Italian MURST under grants cofin-98-2.11.03.07 
%and 99-2.11.03.01 at the University of Padova. 

\newpage\null
\vskip 1.5truecm
\centerline{
\epsfxsize=16.truecm
\centerline{{\epsfbox{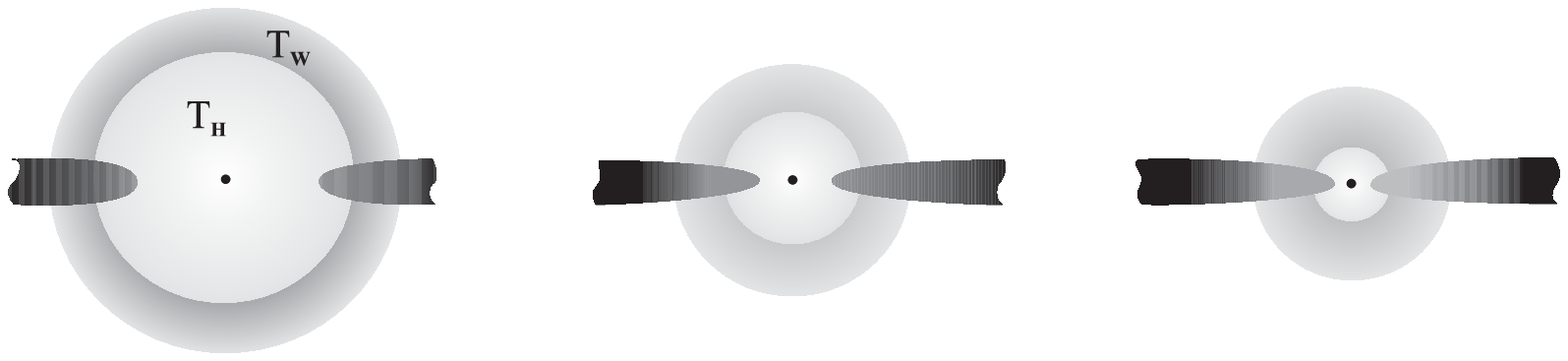}}}}

\bigskip
\figcaption[f1.ps]{A schematic view of the accretion flow in the central region
of GRS 1915+105 in state C. The three panels illustrate the evolution as the 
disk rim 
moves closer to the hole (from left to right). Temperature increases from 
black to white.
\label{artist}}

\vskip 2.truecm 

\centerline{
\epsfxsize=19.truecm
\centerline{{\epsfbox{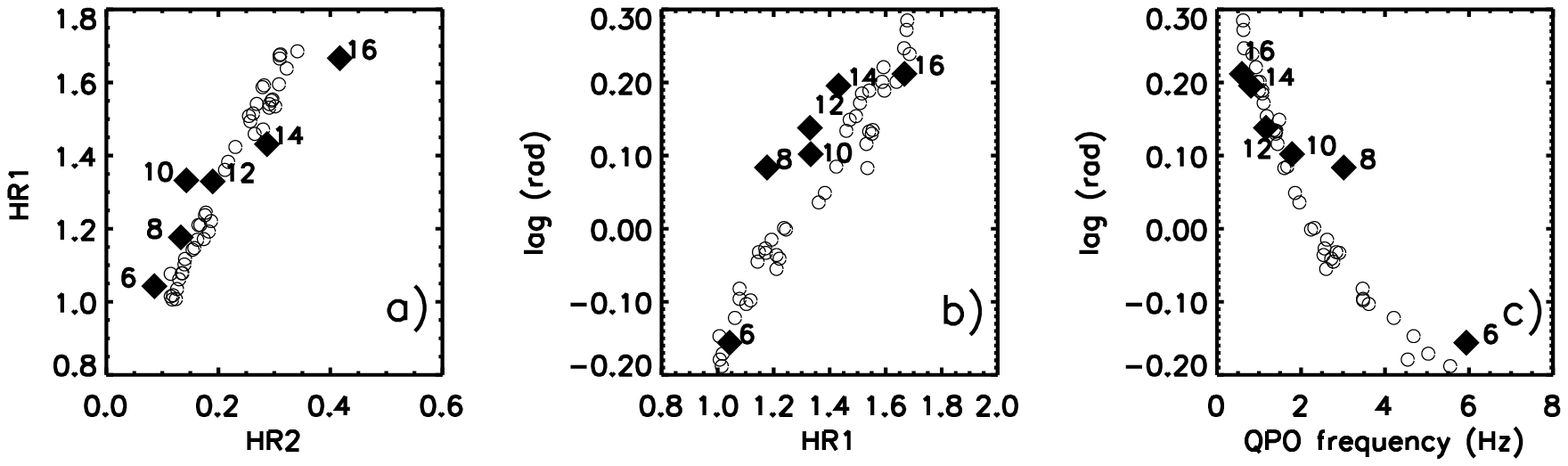}}}}

\figcaption[f2.ps]{a) color-color diagram for the model discussed
in the text
(filled diamonds) compared with observations (open circles); b) predicted and
observed phase
lags of the continuum versus HR1; c)
predicted and observed phase lags of the continuum
versus QPO frequency. Each model is labeled by the value of $r_{in}$ in units
of $GM/c^2$.
\label{plots}}

\end{document}